\documentclass[useAMS,usenatbib]{mn2e}

\title[Intra-night variability in blazars]{The nature of the intra-night optical variability in blazars}

\author[Bachev et al.]
{
R. Bachev$^{1}$, E. Semkov$^{1}$, A. Strigachev$^{1}$,  Alok C. Gupta$^{2}$, Haritma Gaur$^{2,3}$,
\newauthor  B. Mihov$^{1}$, S. Boeva$^{1}$, L. Slavcheva-Mihova$^{1}$\\ 
$^{1}$Institute of Astronomy and National Astronomical Observatory, Bulgarian Academy of Sciences, Sofia 1784, Bulgaria; bachevr@astro.bas.bg\\
$^{2}$Aryabhatta Research Institute of Observational Sciences (ARIES), Manora Peak, Nainital -- 263129, India\\
$^{3}$Department of Physics, DDU Gorakhpur University, Gorakhpur -- 273009, India\\
}

\date{Accepted 2012 May 15. Received 2012 May 11; in original form 2012 March 27}

\pagerange{\pageref{firstpage}--\pageref{lastpage}} \pubyear{2012}

\begin{document}

\maketitle

\label{firstpage}

\begin{abstract}

In this paper we present results of a short-term optical monitoring of 13 blazars. The objects were monitored mostly in the R-band
for a total of $\sim$160 hours between 2006 and 2011. We study the nature of the short-term variations and show that
most of them could be described as slow, smooth, and (almost) linear changes of up to $\sim$0.1 mag/hour, but many objects show 
no short-term variations at all. 
In fact, we found only $\sim 2$ per cent chance to observe variability of more than 0.1 mag/hour 
for the sample we observed.
Hints for quasi-periodic oscillations at very low amplitude levels are also found for some objects.
We briefly discuss some of the possible mechanisms to generate the intra-night variability and the quasi-periodic oscillations.

\end{abstract}


\section{Introduction}

Nowadays it is widely accepted that a large variety of observational blazar phenomena, including the powerful emission, the spectral 
energy distribution (SED) covering practically all energy bands, the exceptional variability, the strong polarization, etc. can be 
attributed to plasma processes in a relativistic jet. According to the commonly invoked scheme, a highly relativistic jet, pointed 
at a small angle with respect to the line of sight (Urry \& Padovani, 1995), generates most of the observed spectrum via synchrotron and inverse 
Compton processes. Due to the relativistic beaming, the emission is highly anisotropic, significantly amplified, and frequency 
shifted with a typical bulk Doppler factor ($\delta$) of around 10--30, as assessed by the SED modeling or variability properties
(Hovatta et al., 2009; Wu et al., 2011b; Rani et al., 2011a).

As the non jet-dominated AGN, generally do not show huge variations, at least not on the intra-night time scales, it is almost certain 
that the blazar variability is also associated with the processes in the relativistic jet (see Wagner \& Witzel, 1995, for a review).
These processes can be intrinsic to the jet and can be related to a change of the jet power, for instance. 
The jet power will change as a result of rapid evolution of the energy 
spectrum of the emitting relativistic particles due to either energy loss (both synchrotron and Compton losses) or energy gain 
(re-acceleration or fresh particle injection), or both at different scales. The processes responsible for the variations can also be 
related to the jet geometry, i.e. a changing jet direction (a curved or swinging jet; Gopal-Krishna \& Wiita, 1992) leading to a variable bulk 
Doppler factor of the blobs traveling along the jet. The last type of processes generates observable variability even if no change of the overall 
jet power occurs. And finally, the variability can be due to entirely extrinsic reasons, like microlensing (Gopal-Krishna \& Subramanian, 1991; 
Paczynski, 1996) from intervening objects (e.g., stars in a foreground galaxy) along the line of sight. Thus, the emission from blobs moving 
relativistically down the jet might vary due to microlensing at a much faster rate than what is normally expected.
Last but not least, the jet itself is not a stationary object; a number of reasons including turbulence, developing/decaying shocks, 
different instabilities, changing environment, etc., can be responsible for the variability, at least at longer time scales.

So, it is not surprising that the exact cause of the blazar variability is still under debates. Solving this problem, however, is of 
importance not only to astrophysics. Understanding blazar variability is also one of the keys to understand the physical processes 
in a relativistic jet. Since the jets are natural accelerators to energies, exceeding on many orders what can be 
achieved on Earth, blazar studies could also be of extreme importance for understanding the fundaments of physics.

This work focuses on the optical variations on the shortest time scales (the so called intra-night or intra-day variability). 
Optical studies are very important for blazars, as even though their spectrum is very broad, a key ingredient of their SED, 
namely the synchrotron peak, is often located close to these wavelengths. 

Although time costly, the intra-night variability studies cannot be replaced by $inter$-night (long-term) monitoring, 
since the short-term magnitude gradients are often much larger than night-to-night gradients (see for instance Fig. 1 in 
Romero et al. 2000; Fig. 2 in Cellone et al. 2007; etc.). Fortunately, most of the objects can successfully be monitored 
with relatively small ($<1$-m) class telescopes, thus making possible to organize dense, multisite monitoring of selected 
objects (e.g., WEBT/GASP campaigns\footnote{Details at http://www.to.astro.it/blazars/WEBT}). 

In this paper we monitor 13 blazars during 58 nights between years 2006 and 2011 for a total of about 160 hours.
%Our goal has been twofold (Sect. 2). First 
Our goal has been to establish the general character of the variations on intra-night time scales. 
Even though there are a number of reports for very rapid, magnitude-scale "frame-to-frame" variations, our observations indicate 
mostly slow trends or wobbles that can successfully be fitted with a low-order polynomial (Sect. 3), and for many objects/nights 
we find no variations at all. 

%Secondly, we want to check statistically (Sect. 4) if any of the existing variability models can be 
%corroborated by the observations. For that purpose we compare the distribution of the magnitude change rates 
%($\mathrm{d}m/\mathrm{d}t$ and $\mathrm{d}^{2}m/\mathrm{d}t^{2}$) for the sample we monitor with the predictions of different models. 

We continue next with describing our observational data.

\section{Observational data}

\begin{table}
 \centering
 %\begin{minipage}{140mm}
  \caption{Observations}
  \begin{tabular}{@{}lrrrrr@{}}
  \hline
\hline

Object	&	UT date	&	Telescope	&	Duration	&	Points & Filter\\%
  
\hline
3C 66A	&	14.10.2007	&	R50/70	&	1.1	&	18	&	R	 \\%
(0219+428)	&	02.11.2007	&	B60	&	2.8	&	79	&	R	 \\%
		&	03.11.2007	&	B60	&	1.9	&	53	&	R	 \\%
		&	08.10.2009	&	A104	&	1.6	&	31	&	R	 \\%
\hline											
1ES 0229+200&	09.10.2009	&	A104	&	5.3	&	67	&	R	 \\%
		&	21.11.2009	&	A104	&	4.4	&	48	&	R	 \\%
\hline											
AO 0235+16	&	20.02.2007	&	R50/70	&	0.9	&	20	&	R	 \\%
		&	25.02.2007	&	R50/70	&	1.0	&	20	&	R	 \\%
\hline											
4C 47.08	&	10.10.2009	&	A104	&	3.8	&	49	&	R	 \\%
(0300+470)	&	22.11.2009	&	A104	&	1.1	&	13	&	R	 \\%
\hline											
S5 0716+714	&	16.01.2007	&	B60	&	1.3	&	19	&	R	 \\%
		&	18.01.2007	&	B60	&	0.9	&	19	&	R	 \\%
		&	24.02.2009	&	B60	&	6.6	&	161	&	I	 \\%
		&	25.02.2009	&	B60	&	6.4	&	180	&	I	 \\%
\hline											
PKS 0735+178&	08.10.2009	&	A104	&	1.6	&	32	&	R	 \\%
		&	20.12.2009	&	A104	&	6.9	&	115	&	R	 \\%
\hline											
PKS 0736+017&	29.02.2008	&	B60	&	5.5	&	77	&	NO	 \\%
		&	03.03.2008	&	B60	&	3.5	&	40	&	NO	 \\%
		&	25.01.2011	&	B60	&	3.3	&	44	&	NO	 \\%
		&	25.04.2011	&	B60	&	0.8	&	21	&	NO	 \\%
\hline											
OJ 287	&	19.11.2006$^{\ast}$	&	B60	&	0.9	&	26	&	R	 \\%
(0851+202)	&	14.12.2006	&	R200	&	4.2	&	98	&	R	 \\%
		&	17.12.2006$^{\ast}$	&	R50/70	&	1.9	&	30	&	R	 \\%
		&	20.02.2007$^{\ast}$	&	R50/70	&	1.9	&	30	&	R	 \\%
		&	08.04.2007$^{\ast}$	&	R200	&	1.8	&	40	&	R	 \\%
		&	09.04.2007	&	R200	&	2.1	&	50	&	R	 \\%
		&	10.04.2007$^{\ast}$	&	R50/70	&	2.9	&	45	&	R	 \\%
		&	29.02.2008$^{\ast}$	&	R50/70	&	2.8	&	75	&	R	 \\%
\hline											
3C 279	&	26.01.2006	&	B60	&	0.5	&	15	&	R	 \\%
(1253-055)	&	27.01.2006	&	B60	&	1.1	&	30	&	R	 \\%
		&	28.01.2006	&	B60	&	0.7	&	20	&	R	 \\%
\hline											
3C 345	&	20.06.2007	&	R200	&	3.4	&	69	&	R	 \\%
\hline											
BL Lac	&	15.07.2007$^{\ast}$	&	R200	&	0.4	&	10	&	R	 \\%
(2200+420)	&	16.07.2007	&	R200	&	3.1	&	47	&	R	 \\%
		&	14.08.2007	&	R200	&	1.9	&	50	&	R	 \\%
		&	15.08.2007	&	R200	&	2.7	&	70	&	R	 \\%
		&	16.08.2007	&	R200	&	2.3	&	75	&	R	 \\%
		&	18.08.2007	&	R50/70	&	3.4	&	70	&	R	 \\%
		&	19.08.2007	&	R50/70	&	3.2	&	58	&	R	 \\%
		&	20.08.2007	&	R50/70	&	3.0	&	55	&	R	 \\%
		&	02.11.2007	&	B60	&	0.7	&	20	&	R	 \\%
		&	07.12.2007$^{\ast}$	&	B60	&	2.0	&	47	&	R	 \\%
		&	08.01.2008	&	B60	&	2.6	&	61	&	R	 \\%
		&	09.01.2008$^{\ast}$	&	B60	&	2.3	&	58	&	R	 \\%
		&	11.01.2008	&	B60	&	1.8	&	49	&	R	 \\%
		&	07.07.2008	&	R200	&	1.5	&	74	&	R	 \\%
		&	08.10.2009	&	A104	&	4.0	&	78	&	R	 \\%
\hline											
3C 454.3	&	28.07.2008	&	B60	&	1.9	&	54	&	R	 \\%
(2251+158)	&	29.07.2008	&	B60	&	2.2	&	62	&	R	 \\%
		&	30.07.2008$^{\ast}$	&	B60	&	2.2	&	63	&	R	 \\%
		&	31.10.2010$^{\ast}$	&	R50/70	&	5.0	&	27	&	R	 \\%
		&	01.11.2010$^{\ast}$	&	R50/70	&	5.8	&	27	&	R	 \\%
		&	02.11.2010$^{\ast}$	&	R50/70	&	5.3	&	30	&	R	 \\%
		&	03.11.2010$^{\ast}$	&	B60	&	6.7	&	57	&	R	 \\%
		&	04.11.2010$^{\ast}$	&	B60	&	3.1	&	29	&	R	 \\%
		&	06.11.2010$^{\ast}$	&	R50/70	&	3.6	&	21	&	R	 \\%
\hline											
B2 2308+341	&	10.08.2010	&	B60	&	2.8	&	40	&	NO	 \\%
		&	11.08.2010	&	B60	&	2.1	&	30	&	NO	 \\%

\hline
\hline
\end{tabular}
%\end{minipage}
\end{table}

We employed several instruments equipped with CCD cameras and standard UBVRI filter sets to perform the photometric monitoring. 
These include the 60cm reflector of Belogradchik Observatory, Bulgaria, (coded B60 in Table 1), equipped with SBIG ST-8 CCD 
(as of 2009 replaced with FLI PL9000); the 2m RCC reflector (Photometrics AT200 CCD) and the 50/70cm Schmidt camera (SBIG ST-8, later 
replaced with FLI PL16803 CCD) of Rozhen National Observatory, Bulgaria, (coded R50/70 and R200 respectively); and the 104cm reflector of 
ARIES, Nainital, India (Wright 2k CCD), coded as A104. 
Further details on the instruments in use can be found in Gaur et al. (2012b).

A total of almost 3000 individual frames were obtained and analyzed\footnote{Some of these data may have been also used in 
active WEBT campaigns}. The typical exposure time was 120s; however, depending on 
the brightness of the object and the telescope size, different exposure times between 60 and 240s were used on different occasions. 
The monitoring was performed mostly in the R-band with a few exceptions, where the I-band or no filter were used (Table 1). The latter was 
the case for two very weak objects, monitored with a relatively small telescope; see Bachev et al. (2005) for comments on this issue. 
The majority of the nights were clear and stable but on a few occasions we monitored during non favorable atmospheric conditions, 
which resulted in larger and variable photometric errors (see Sect. 4.1 for discussions).

After bias, dark current (where appropriate) and flat field corrections, the magnitudes were extracted by applying standard aperture 
photometry routines. The aperture radius was taken to be typically 2--3 times the seeing. The magnitudes of the blazar and a check star 
were measured with respect to a main standard of known magnitude with no further corrections made. Both, the main standard and the 
check star were chosen on the basis of their magnitude, color, and spatial proximity to the blazar;
however, due to the different instrument characteristics (fields of view), no formal criteria were applied.

Table 1 gives a short log of the observations. The columns give the object name, the UT date when the observation started, the instrument, 
the monitoring duration in hours, the number of the observational points for the corresponding run, and the filter. Asterisk after the date 
indicates an observation, used for the statistics but not shown in the figures.

\begin{figure}
 %\mbox{} \vspace{8.5cm} \special{psfile=3C66.eps hoffset=-20 voffset=-320
\mbox{} \vspace{6.0cm} \includegraphics{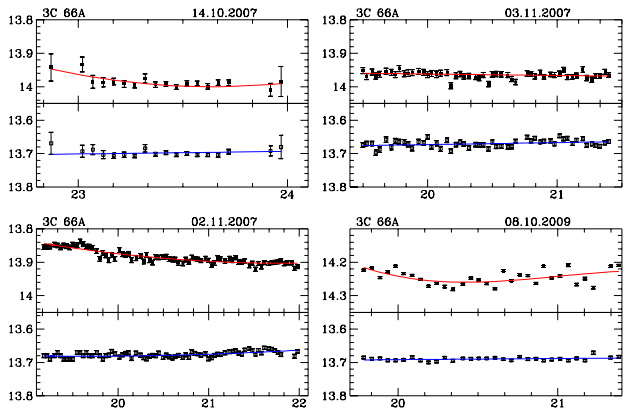} 

\caption[]{
Intra-night blazar LCs (3C 66A data). The upper panel of each 
box shows the blazar LC (filled symbols), measured with respect to a main standard, and the lower one -- the same for a 
check star (open symbols). The photometric errors at 1-$\sigma$ level are indicated as error bars.
The best fitting low-level polynomial (see the text) is also shown for both -- the blazar and the check star LCs. 
The name of each monitored object and the UT date when the observation started are shown atop of each box. 
The abscissa shows the UT in hours. Each upper panel has the same vertical scale factor as the corresponding lower panel, 
but the exact value may differ from object to object.
}
\end{figure}

\begin{figure}
\mbox{} \vspace{6.0cm} \includegraphics{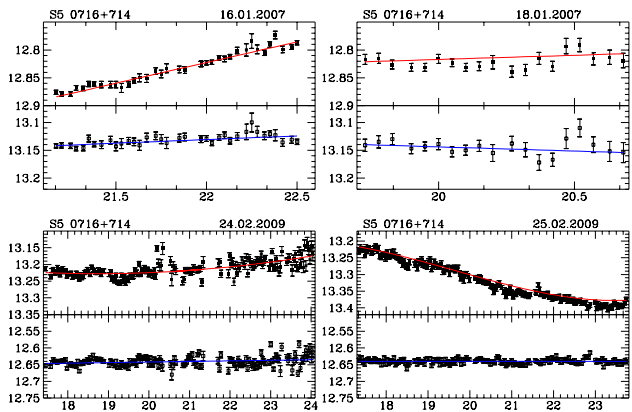} 

\caption[]{
The same as Fig. 1 (S5 0716+714 data)
}
\end{figure}

\begin{figure}
\mbox{} \vspace{6.0cm} \includegraphics{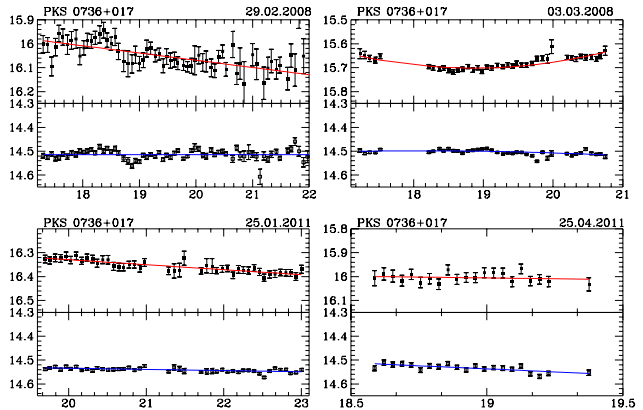} 

\caption[]{
The same as Fig. 1 (PKS 0736+017 data)
}
\end{figure}

\begin{figure}
\mbox{} \vspace{8.5cm} \includegraphics{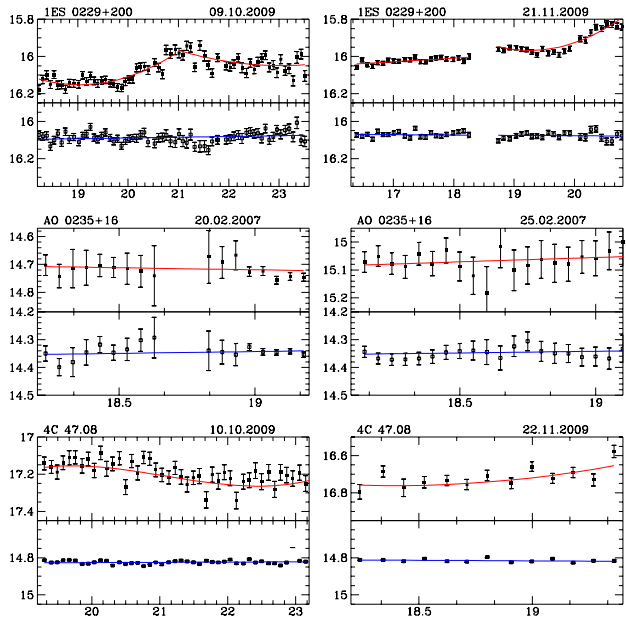} 

\caption[]{
The same as Fig. 1 (1ES 0229+200, AO 0235+16, and 4C 47.08)
}
\end{figure}

\begin{figure}
\mbox{} \vspace{8.5cm} \includegraphics{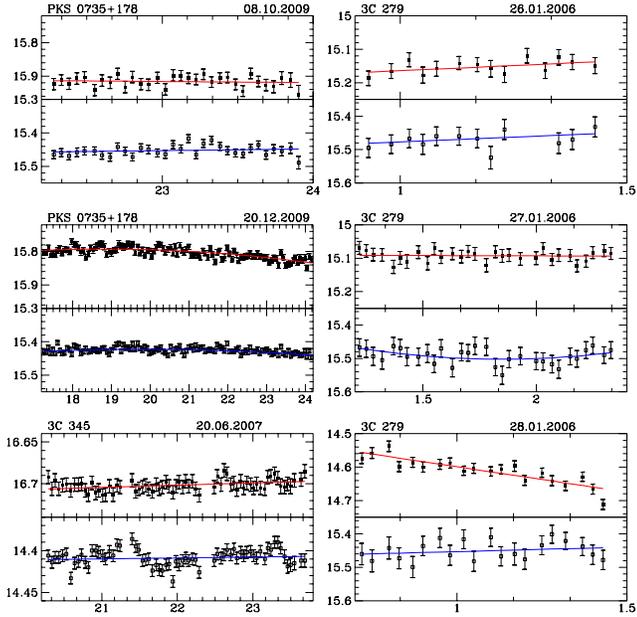} 

\caption[]{
The same as Fig. 1 (PKS 0735+178, 3C 279, and 3C 345)
}
\end{figure}

\begin{figure}
\mbox{} \vspace{8.5cm} \includegraphics{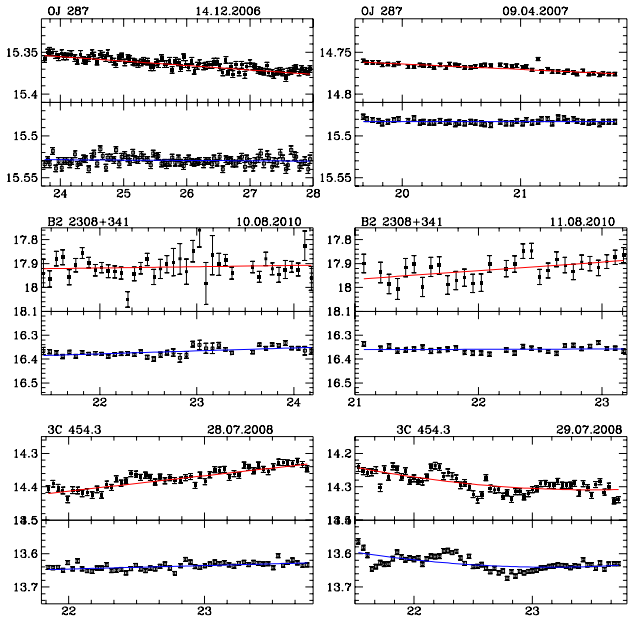} 

\caption[]{
The same as Fig. 1 (OJ 287, 3C 454.3 and B2 2308+341). For OJ 287 only 2 out of 8 LCs are shown. For 3C 454.3 only 
2 out of 9 LCs, are shown; see also Bachev et al., 2011, for the rest of them.
}
\end{figure}

\begin{figure}
\mbox{} \vspace{8.5cm} \includegraphics{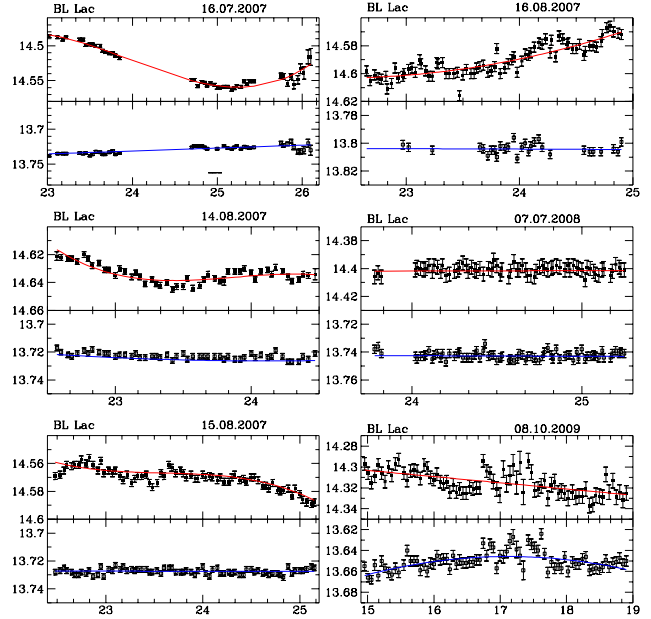} 

\caption[]{
The same as Fig. 1 (BL Lac, larger telescopes data, 6 out of 7 nights shown)
}
\end{figure}

\begin{figure}
\mbox{} \vspace{8.5cm} \includegraphics{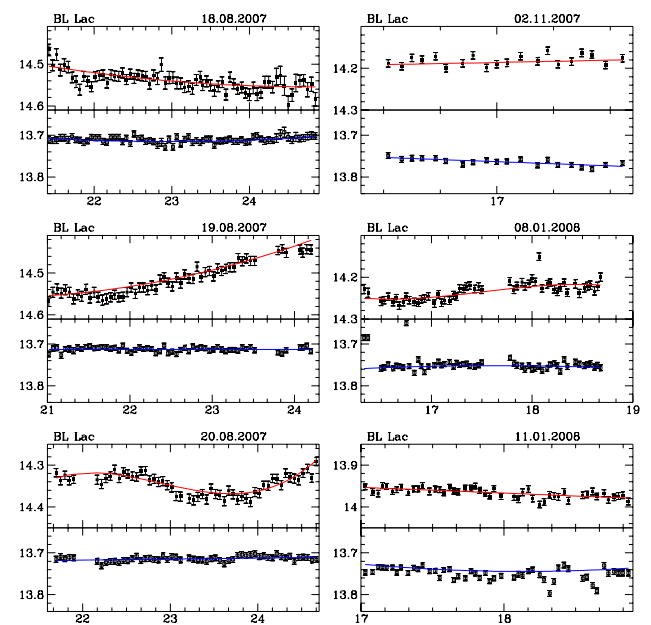} 

\caption[]{
The same as Fig. 1 (BL Lac, smaller telescopes data, 6 out of 8 LCs shown)
}
\end{figure}

\section{Results}

Figs. 1 -- 8 show the results of our intra-night variability study. The top panel of each box shows the blazar 
light curve (LC) and the bottom one -- the check star LCs, both measured with respect to the main standard. 
The object name and the UT date are given atop of each box.

\subsection{The nature of the intra-night variations}

As one can see from Figs. 1 -- 8, the blazar LCs can be described as smooth fluctuations or trends without any violent, 
"frame-to-frame" changes. Furthermore, no statistically significant variations are present at all on many occasions. 
Therefore, it would be appropriate to fit the LCs with a straight line or -- less often ($\sim$40\% of the cases) -- with a low-order 
polynomial. This approach enables us to better represent the variations and to eliminate to a large extent any spurious effects introduced 
by the photometric errors. Fitting a low order polynomial was already successfully applied by Bachev et al. (2011) to search for time 
delays between different optical bands (see also Montagni et al. 2006, who fitted straight lines to fit different segments of the LCs). 
The corresponding best fits are also shown in Figs. 1--8 for the LCs of both -- blazars and check stars. As one sees, the fits are rather successful 
in most (but perhaps not all!) cases even though some of the runs lasted for more than 5 hours.
Fitting LCs allows studying the distribution of magnitude change rates, $\mathrm{d}m/\mathrm{d}t$,
%and $\mathrm{d}^{2}m/\mathrm{d}t^{2}$
for all objects individually, as well as for the entire sample (Sect. 3.3).

The order of the polynomial was chosen to be as low as possible and a linear slope was used in almost $\sim50\%$ of cases. 
Unfortunately, there is no easy way to use statistics (e.g., $\chi^{2}$) to determine the degree of the polynomial. 
The reason is that the real photometric errors are often larger (e.g., Bachev et al., 2005, and the references therein) 
than the theoretical ones by a factor that is difficult to determine and may well vary from instrument to instrument and/or 
observing conditions. Furthermore, correcting the theoretical errors by a certain factor does not seem to be a good practice 
(de Diego, 2010). We have to mention, though, that the distributions of the magnitude 
change rates depend very little on the exact polynomial order, as long as this order is kept much lower than the number of data points.

Even though we find no rapid variations (say more than a few hundreds of a magnitude on a minute time scale), 
there are sometimes such claims in the literature, some of which concern objects from our sample. It is difficult to judge on the 
reality of these results as often there is no independent way to verify them. See Cellone et al. (2007) for examples and a critical analysis.

\subsection{Individual notes}

\noindent {\bf 3C 66A}. 
During 4 nights of monitoring (Fig. 1), this TeV blazar showed only minor ($\sim 0.05$~mag/hour) variations at best. This behavior 
is consistent with the reports by Raiteri et al. (1998), Dai et al. (2001), Sagar et al. (2004), B\"{o}ttcher et al. (2009), Rani et al. 
(2011b), where 3C~66A was observed at a similar brightness stage and showed similar short-term variability. 
Although a rapid change was reported on one occasion ($\Delta\rm B\simeq0.5$ mag/hour; Zhang et al., 2004), overall, the object does 
not seem to be highly active most of the time, showing only minor, smooth variations.

\noindent {\bf 1ES 0229+200}. 
We monitored this VHE $\gamma$-ray source for 2 nights for a total of $\sim 10$ hours (Fig. 4). To the best of our knowledge, 
% this object has never been optically monitored before on intra-night time scales, 
no results of previous intra-night optical monitoring of this object were published,
and Kurtanidze et al. (2004) found no variations on longer time scales. 
Our monitoring, however, indicates the presence of rapid intra-night variability of up to 0.15 mag/hour. As a matter of 
fact, this is one of the few objects, where even a 3th degree polynomial did not seem to fit the LC very well.
For this particular object we had to use two different polynomials to fit the entire LC.

\noindent {\bf AO 0235+164}.
We monitored this object as a part of an active WEBT campaign (Raiteri et al., 2008a) during 2 photometrically poor nights 
(Fig. 4) and found practically no intra-night variations, but a change of $\Delta\rm R\simeq0.4$ mag was seen within $\sim5$ days.
Other authors (Hagen-Thorn et al., 2008), however, report significant intra-night variations of $\Delta\rm R\simeq 0.5$ mag 
within several hours. Romero et al. (2000, 2002) also observed rapid changes, indicating that variations of $\Delta\rm R\simeq 0.1$ mag/hour 
may be typical for this otherwise highly active object (see also Sagar et al., 2004; Gupta et al., 2008; Rani et al., 2011b).

\noindent {\bf 4C 47.08}. 
This \textit{Fermi} LAT source (Abdo et al., 2010) has been rarely studied at optical wavelengths. We found a $\sim 0.5$ mag 
variation on a month time scale, but only minor variations during the 2 nights of intra-night monitoring. The object was rather weak 
and the photometric uncertainties might have contributed to the apparent variations, seen as smooth fluctuations (Fig. 4).
We note that Star C5 (the closest standard, Fiorucci et al., 1998) appears to be variable on short time scales.

\noindent {\bf S5 0716+714}. 
This blazar has been a target of a number of intra-night monitoring campaigns. Practically all studies, e.g., Sagar et al. (1999), 
Nesci et al. (2002), Wu et al. (2005), Gu et al. (2006), Montagni et al. (2006), Pollock et al. (2007), Gupta et al. (2008), 
Stalin et al. (2009), Poon et al. (2009), Carini et al. (2011), Rani et al. (2011b), 
report significant variations, which can typically be described as smooth fluctuations or a sequence of trends, often abruptly changing direction. 
Some reports even imply the presence of very rapid (tenths of a magnitude) changes within a few minutes (Fan et al., 2011) 
and/or quasi-periodic oscillations (Gupta et al., 2009; Rani et al., 2010) on $\sim15$ min time scale.
Clearly, S5~0716+714 is one of the most active blazars in terms of rapid intra-night variations.
Our monitoring during 4 nights showed smooth trends, reaching up to $\sim0.05$ mag/hour but no rapid changes on a minute time scale (Fig. 2)

\noindent {\bf PKS 0735+178}. 
This is another object very extensively studied in the optical wavelengths throughout the years. Some of the more recent studies by 
Bai et al. (1999), Sagar et al. (2004), Stalin et al. (2004), Gupta et al. (2008), Goyal et al. 
(2009), Rani et al. (2011b) report only low-level intra-night variations, if any at all.  
On the other hand, Zhang et al. (2004) find variability of $\sim0.5$ mag/hour in one occasion. Our observations 
confirm what most of the researchers report, i.e. we found practically no variability during 2 nights for a total of about 
8.5 hours of monitoring (Fig. 5).

\noindent {\bf PKS 0736+017}. 
This FSRQ is known to show occasionally exceptional variability on intra-night scales. Clements et al. (2003), for instance, reported a 1.3 
mag outburst within 2 hours. Their observations, however, as well as our results imply that PKS 0736+017 is much more active, 
when close to its maximum (R$\simeq$13.7 mag), or at least that periods of intra-night activity are followed by periods of  
almost constant brightness. Sagar et al. (2004) also reported very quiescent behavior when the object was around R$\simeq$15 mag. 
Our observations (Fig. 3) were during a deep minimum -- $\sim$16 mag (see also Ram´irez et al., 2004), yet some low-level, 
smooth variations are seen.

\noindent {\bf OJ 287}. 
This object manifested very little if any intra-night variability during 8 nights of monitoring, spanning over $\sim$ 15 months, 
though night-to-night variations of $\sim$0.1 mag were present. In Fig. 6 we only show the results from 2 atmospherically 
stable nights, where gradual trends are clearly seen. In spite that we found no rapid variations during our monitoring of this object, 
other authors report significant changes on very short time-scales. Zhang et al. (2007), for instance, found a $\Delta R\simeq 0.35$ mag 
change for a half an hour during one night (see also Xie et al., 2001). On the other hand, Carini et al. (1992), 
Gonzalez-Perez et al. (1996), Dultzin-Hacyan et al. (1997), Sagar et al. (2004), Gupta et al. (2008), Rani et al. (2011b),
report only minor variations, if any, which is much more consistent with our findings.

\noindent {\bf 3C 279}. 
The object was monitored in 3 consecutive nights in Jan. 2006, being around 15 mag (much fainter than its historical maximum, 
R$\simeq 12.5$ mag). A clear decreasing trend of 0.15 mag/hour, probably tracing the end of a small outburst 
(Fig. 5, see also B\"{o}ttcher et al., 2007, for a long-term LC covering this period) was seen in one night, while during the other two 
nights the object was quite stable. Kartaltepe \& Balonek (2007) reported similar gradual trends (if any) during their monitoring, 
when the object was around 14 mag. Gupta et al. (2008) found practically no variability during their monitoring, but on the other hand, 
rapid changes were reported by Miller et al. (2011) for low state. Clearly, 3C~279 exhibits different types of 
intra-night variability behavior.

\noindent {\bf 3C 345}
This superluminal FSRQ has not been observed for intra-night variability very often, probably because it has been rather weak 
($\sim 16-17$ mag) in the recent years. Mihov et al. (2008) and Howard et al. (2004) found practically 
no intra-night variability. On the other hand, Wu et al. (2011a) reported trends of $\sim0.2$ mag/hour, while 
Kidger \& de Diego (1990) found even sharper decline in one occasion. We monitored this object during one 
night and found no variations at all (Fig. 5).

\noindent {\bf BL Lac}.
We monitored extensively the blazar archetype for a total of almost 35 hours during 15 nights, in 6 of which a 2-meter class 
telescope was used (Figs. 7, 8). Variability pictures reveal trends of up to $\sim$0.1 mag/hour or smooth fluctuations. The object 
was mostly around $R\sim14.5$ mag, meaning a rather low state, and showed little night-to-night changes (Raiteri et al., 2009). 
Other researchers report similar intra-night behavior (Nesci et al. 1998; Speziali \& Natali 1998; Papadakis et al. 2003; 
Howard et al. 2004), e.g., smooth variations of typically $\sim$0.1 mag/hour. On the other hand, Zhang et al. (2004) found 
unusual $\simeq$0.3 mag frame-to-frame variations.

\noindent {\bf 3C 454.3}
This is another well studied FSRQ. Some of the recent intra-night variability studies include Poggiani (2006), 
Gupta et al. (2008), Rani et al. (2011b), Gaur et al. (2012a) who found only minor or no variations, 
but $\sim 0.2$ mag/hour variability rates were seen occasionally. Similar behavior (see also Fig. 6) was also reported by 
Bachev et al. (2011), who monitored the object during a recent outburst to search 
for time delays between different optical bands on intra-night time scales. Their R-band LCs were also used for the purposes 
of this work. Zhai et al. (2011) performed a similar search but none of 
these works found convincing evidence for a delay between the optical bands.
Raiteri et al. (2008b), on the other hand, give clear examples of violent intra-night variability episodes of 
$\sim 1$ mag/hour perhaps tracing short-lasting outbursts during the high activity phase of the blazar. Obviously, this object 
demonstrates different variability behavior, probably depending on its brightness state.

\noindent {\bf B2 2308+341}
Optical monitoring of this object was triggered by increased $\gamma$-ray activity (D'Ammando, 2010). 
The object was very weak, so no filter was used during the 2 nights of monitoring. Thus, the magnitudes shown in Fig. 6 are 
somewhat arbitrary. No huge intra-night or day-to-day variations were observed. The object has been rarely studied and we 
found no other works on intra-night variability.

\begin{figure}
\mbox{} \vspace{7.0cm} \includegraphics{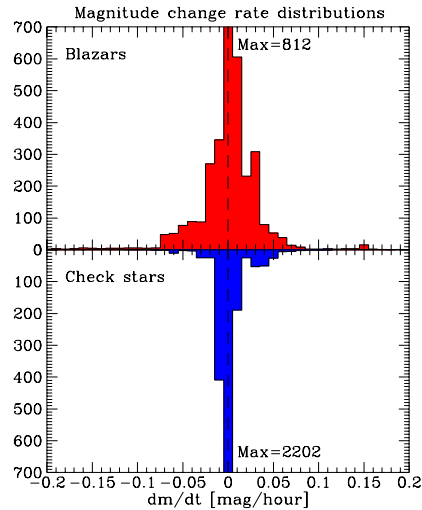} 

\caption[]{
Magnitude change rates for the blazars and the check stars, as found from the fitting polynomials.
The numbers in the distribution count the 3 min intervals from the LC with the corresponding magnitude change rate.
}
\end{figure}

\subsection{Distribution of the variability rates of the sample}

The LCs we analyze in this work cover a total of $\sim 160$ hours of monitoring. Although the LC data can not be considered 
complete nor can the sample be considered representative, it still may be worth building a histogram of the magnitude change rates 
($\mathrm{d}m/\mathrm{d}t$) for the entire sample. Such a histogram may help studying time asymmetries, for instance.
Fig. 9 shows the distribution of $\mathrm{d}m/\mathrm{d}t$ values for the blazars (upper panel) and the check stars (lower panel) based on the 
corresponding best fits. Normally, one would not expect the check stars to show variations; however, sometimes slow trends 
became evident after applying the same fitting procedure as for the blazars, resulting in some non-zero values in the $\mathrm{d}m/\mathrm{d}t$ 
histogram for the check stars. Explanations can be searched for in terms of variable atmospheric conditions; possible low-level stellar 
variability; etc. (see Klimek et al., 2004; Bachev et al., 2005), but this effort goes beyond the scope of this paper.

There are two important results that are worth mentioning. First, it seems that observing significant variability rates is a very rare case.
For instance, from Fig. 9 we find only $\sim 2\%$ chance to observe blazar variability $|\mathrm{d}m/\mathrm{d}t|>0.1$ mag/hour and $\sim 25\%$ chance 
to observe practically no detectable variability, $|\mathrm{d}m/\mathrm{d}t|<0.005$ mag/hour (the corresponding percentage for the check stars is $\sim 68$).
Secondly, the distribution does not appear to be symmetric. Statistical tests to compare the means, medians, standard deviations, 
as well as the KS test, performed separately on the "negative" and "positive" parts of the $\mathrm{d}m/\mathrm{d}t$ distribution distinguish them at 
99.9\% level, even though the mean of the entire distribution is consistent with zero within the errors (as expected). This might be 
a result of the data incompleteness; however, it also might be an artifact of a real time asymmetry of the blazar LCs.
Since the total monitoring times were very different for different objects, we cannot claim that such asymmetry, if real at all, 
is a general property of the sample or, at even lesser extent, of the entire blazar population.

\subsection{Possible quasi-periodic oscillations}

Employing a 2-m class telescope to monitor some of the objects (BL Lac in particular) ensured photometry of very high accuracy 
(e.g., 0.002 -- 0.003 mag) during some stable nights, which on the other hand allowed studying possible (quasi-periodic) micro-oscillations. 
Such have been previously reported; e.g., Clements et. al. (2003) for PKS 0736+017; Wagner et al. (1996), Gupta et al. (2009), 
Rani et al (2010) for S5 0716+714; etc. 
For BL Lac, some hints for quasi-periodicity were presented by Speziali \& Natali (1998, their Fig. 3), though the authors do not claim 
any statistical significance. It is clear from our BL Lac data (the most extensively studied object in our sample) that a 
periodic signal is not the dominating feature in the LCs. However, after subtracting the leading polynomial, one may attempt to search 
for micro-oscillations in the residuals. We found some clues for the presence of such micro-oscillations with an amplitude of $\leq0.01$ mag 
and a typical period $T\sim$1 hour for many of the nights (Fig. 10).

Interestingly, the oscillation pattern does not appear to depend much on the order of the leading polynomial (within some limits, of course), 
used to fit the LC. If further confirmed, micro-oscillations with a period of an hour or so may have significant implications for our 
understanding of how matter accelerates along the jet. We note on the other hand, that such a period is close to the shortest 
possible \textit{Keplerian} time (perhaps even below the innermost stable orbit, Sect. 4.3).

\begin{figure}
 \mbox{} \vspace{6.0cm} \includegraphics{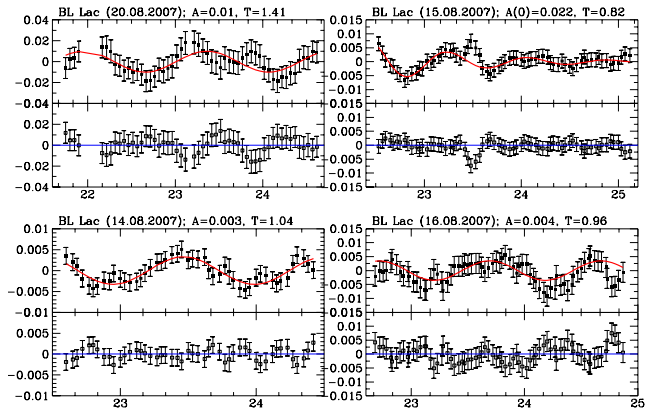} 
\caption[]{BL Lac light curves after subtracting the leading polynomial (Fig. 7). Moving average filter is applied to the
data to reduce the scatter. Micro-oscillations are clearly seen in some nights (e.g., 14.08.2007). The upper panels show the 
sine function best fits, and the lower ones -- the residuals. For the night of 15.08.2007, an exponentionally decaying 
sine amplitude works best.
}
\end{figure}

\section{Discussion}

\subsection{Differential light curves and atmospheric stability}

Since the colors of the blazars and the stars used for the differential photometry may sometimes differ significantly, 
one can raise the question about the influence of the atmospheric stability and/or air mass change on the differential LCs. 
In order to check for such effects we consider closely cases when the blazar showed significant variations during the 
monitoring in unstable nights. Fig. 11 shows two such cases. The upper boxes show the blazars' LCs and the lower 
ones -- the signal from the main standard, transformed into arbitrary magnitudes. One sees that while the main standard's 
magnitude varied significantly, both because of the changing air mass and the unstable atmosphere, the differential blazar 
LCs do not seem to be influenced at all, which leads to the conclusion that the observed trends are unlikely to be attributed 
to the changing atmospheric transparency. The Pearson correlation coefficients between the two datasets are $-0.36$ for S5 0716+714 
and $-0.66$ for PKS 0736+017, implying a rather insignificant correlation. Spearman rank correlation coefficients are even 
less significant -- $-0.27$ and $-0.41$, respectively.
As a matter of fact, a small anticorrelation between the datasets is actually expected, as the photometric errors of 
the main standard influence both datasets.

\begin{figure}
 \mbox{} \vspace{3.0cm} \includegraphics{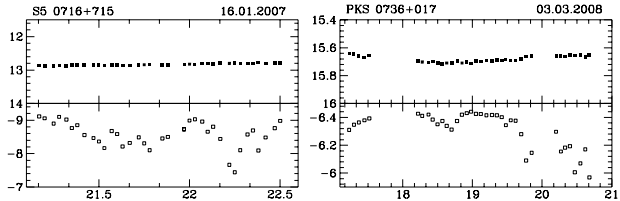} 
\caption[]{The influence of the atmospheric transparency on the differential light curves of
the blazars -- S5 0716+714 and PKS 0736+017 monitored during two unstable nights.
The upper boxes show the blazars' LCs and the lower ones -- the signal from the main
standard, transformed into arbitrary magnitudes. One sees that the atmospheric stability does 
not have any visible effect on the differential LCs.}
\end{figure}

\subsection{Implications for the variability models}

\subsubsection{Time asymmetry of the LCs}

Studying the magnitude gradients' distribution may help to disentangle among different variability scenarios. 
For instance, a dissipative process (i.e. an explosive event, etc.), being by nature not time reversible, will likely 
produce a time-asymmetric LC (perhaps with sharp raises and slow declines), which will result in different 
positive and negative magnitude change rate distributions. On the other hand, a process described by a conservative 
dynamical system, whose \textit{Hamiltonian} is not directly depending on time\footnote{See however for more details, 
e.g., Lamb \& Roberts, 1998, and the references therein} (like microlensing, orbital motion, etc.) will 
likely produce a time-symmetric LCs, leading to a symmetric positive/negative gradient's distribution. Unfortunately, our data (Fig. 9) 
cannot be considered statistically significant for a number of reasons (Sect 3.3) for a firm conclusion.

\subsubsection{Evolution of the electrons' energy density distribution}

A natural cause of the brightness changes can be the evolution of the relativistic particles' energy density distribution, $n(\gamma)$, 
which consequently leads to a variable synchrotron emission. While the energy loss rates are known (both synchrotron and 
Compton losses are $\propto\gamma^{2}$, $\gamma$ is the electron's Lorentz factor; Rybicki \& Lightman, 1979), the energy gain rates 
(the acceleration/injection mechanisms) are generally unknown. This leads to the impossibility to solve the corresponding 
\textit{Fokker-Planck} equation (Rybicki \& Lightman, 1979) without a priori imposed assumptions. 
Electron energy density evolution will probably lead to time delays between different wavebands (Bachev et al., 2011, 
and the references therein). Knowing that our LCs are monochromatic and taking into account the above said, places 
testing this variability mechanism beyond the scope of this paper.

\subsubsection{Swinging jet}

\begin{figure}
 \mbox{} \vspace{6.0cm} \includegraphics{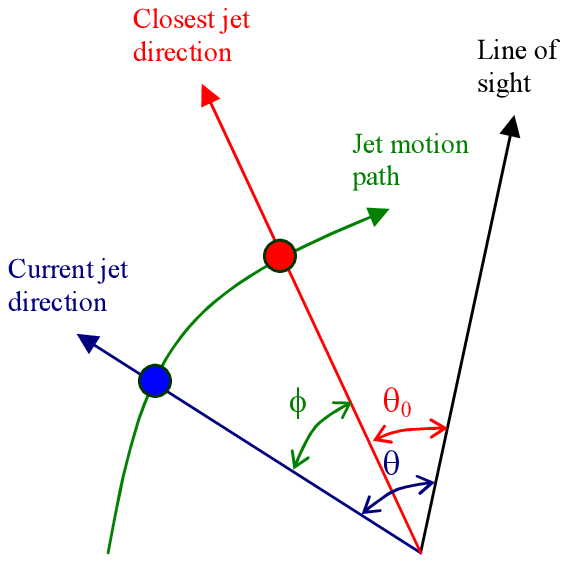} 
\caption[]{
A sketch showing the change of the jet direction and the corresponding angles (see the text).
}
\end{figure}

The idea behind this variability mechanism is that the path of the relativistically moving blobs along the jet can deviate 
slightly from a straight line, thus leading to a variable Doppler factor and respectively to variable emission from that
particular blob (Fig. 12 ilustrates the situation, see also Gopal-Krishna \& Wiita, 1992). 
From the definition of the Doppler factor $\delta \equiv \frac{1}{\Gamma(1-\beta \cos\theta)}=\frac{1}{\Gamma(1-\beta\mu_{\rm 0}\cos\varphi)}$, 
where $\mu_{\rm 0}=\cos\theta_{\rm 0}$ and $\theta_{\rm 0}$ is the "impact parameter", i.e. the smallest angle between the jet direction 
(direction of motion of the blob) and the line of sight. Clearly, the maximum amplification occurs when $\theta_{\rm 0}=0$. 

%Let the jet direction changes with a constant angular speed, $\mathrm{d}\varphi'/\mathrm{d}t'=\Omega'$ 
%(here and below primes indicate rest frame measurements). 
%Taking into account that $\Omega'\simeq\Omega$ (small angle approximation) leads to $\varphi=\Omega t$ 
%if for simplicity we take $t_{(\varphi=0)}=0$. 

If take into account that all angles are small (we consider only cases where the magnification is significant), it can be shown that 
$\delta=\frac{A}{1+ (t/\tau)^{2}}$, where $A$ is a constant and $\tau$ is a characteristic timescale. 

%Here $A=\frac{1}{\Gamma(1-\beta\mu_{\rm 0})}$, which is $\equiv\delta_{\rm max}$ 
%for $\mu_{\rm 0}=1$; and $\tau=\frac{\sqrt{2(1-\beta\mu_{\rm 0})}}{\Omega}$, which is $\simeq \Gamma/\Omega$ for $\mu_{\rm 0}=1$ (as expected).

The magnitude change associated (if this blob is dominating the blazar emission at some particular moment) can be calculated 
taking into account that the flux $F_{\rm \nu}=F_{\rm \nu'}\delta^{3+\alpha}$, where $\alpha$ is the spectral index, i.e.,
$F_{\rm \nu}\propto\nu^{-\alpha}$. Thus, $\mathrm{d}m/\mathrm{d}t \simeq \mathrm{d}\ln\delta/\mathrm{d}t$ (see also Nesci et al., 2002; Montagni et al., 2006). Knowing $\mathrm{d}m/\mathrm{d}t$ 
one can integrate to obtain the magnitude of the blob emission as a function of time.

% (Fig. 13), as well as its second derivative (Sect. 4.2.5).

As expected, $m(t)$ is in general a complex peak function, but for practical purposes it can be approximated with a simpler one. 
For $t>>\tau$, the peak appears quite sharp and the best working simple function here is the exponent. However, the magnitude 
change becomes significant; 
%it is $\sim 3$ mag for $t=\tau$ ($\alpha=1$) and 
e.g., $\sim 10$ mag 
%for $t=3\tau$ 
(see also Gopal-Krishna \& Wiita, 1992). 
Since such large amplitudes are never observed in the blazar LCs, one has to assume that a flare from a single blob, due to a 
variable Doppler factor, may never completely dominate the blazar emission, except perhaps for $t\leq\tau$. Thus, a correct 
approach would be to invoke an additional constant flux component of magnitude $m_{\rm 1}$, which can be due to the host galaxy emission 
and/or to an ensemble of independent flaring events, each at different evolutionary state (but yet, the most of them quite far from the maximum).

Introducing the constant component changes significantly the shape of the flares; this time the best working fit appears to be a \textit{Gaussian} 
for a fairly broad range of ratios between $m$ and $m_{\rm 1}$. In other words, unless we accept that a flare can be so powerful to 
increase the total blazar emission by many orders (say by more than 10 mag), its overall shape has to resemble a Gaussian within the 
framework of this model.

%The Gaussian flare can be presented as $m(t)=m_{\rm 1}-\Delta m \exp(-(t/t_{\rm G})^{2})$, where $\Delta m$ is the amplitude of the combined 
%magnitude and $t_{\rm G}$ is a characteristic time, for which we found $t_{\rm G}/\tau \simeq 0.0048\Delta m^{2}+0.081\Delta m +0.55$, 
%i.e. $t_{\rm G}\sim \tau$ within some limits for a broad range of combined amplitudes ($\sim0.1 - 10$ mag.).

%\begin{figure}
% \mbox{} \vspace{7.0cm} \special{psfile=fig_LCn.eps hoffset=-20 voffset=-420
% hscale=60 vscale=60} 
%\caption[]{
%Typical flare shapes for the swinging jet (upper box) and the microlensing models (lower box). All flares start from the maximum 
%($t=0$) as the LCs are time-symmetric. The actual LCs for the models are given in color and the corresponding fits 
%(see the text) -- in black. For the swinging jet model the amplitude of the flare is controlled mostly by $m_{\rm 1}$ 
%(see the text). Vertical offsets to the LCs may have been applied for presentation purposes. For the microlensing scenario the 
%impact parameters are $\hat\theta_{\rm 0}=0.1$ and 0.5 for the upper and the lower LCs, respectively, with some arbitrary $m_{\rm 1}$.
%}
%\end{figure}

\subsubsection{Microlensing}

We can use a similar scheme to compute the LC of an emitting blob, which emission is magnified as a result of microlensing 
from a star (a point-mass lens) along the line of sight (Paczynski, 1996). 
%Let us assume that the transverse motion of the blob is along a straight 
%line with a constant angular velocity $\mathrm{d}\varphi/\mathrm{d}t=\Omega$. Here the unknown constant $\Omega$ depends on the actual angular velocity 
%with the relativistic corrections, as $t$ is measured in the laboratory system. Unlike the previous case, this time we can 
%assume that $\delta$ is a constant. The minimum angle between the source and the lens, the "impact parameter" is 
%$\theta_{\rm 0}$ and is close to zero to achieve observable magnification. The full angle ($\theta$) between the source and 
%the lens at any given moment can be found from the relation 
%$\theta^{2}=\theta_{\rm 0}^{2} + \varphi^{2} = \theta_{\rm 0}^{2} + (\Omega t)^2 $
%(small angle approximation; $t=0$ and $\varphi=0$ for $\theta=\theta_{\rm 0}$). 
The flux magnification for gravitational lensing is
$M=\frac{\hat{\theta}^{2}+2}{\hat{\theta}\sqrt{\hat{\theta}^{2}+4)}}$, 
where $\hat{\theta}=\theta/\theta_{\rm E}$, where $\theta_{\rm E}$ is the Einstein radius, 
(see for instance Narayan \& Bartelmann, 1996, for a review). 

%Therefore 
%$M(\tau)=\frac{2+\hat{\theta}_{\rm 0}^{2}+\tau^{2}}{\sqrt{(\hat\theta_{\rm 0}^{2}+\tau^{2})^{2}+4\hat{\theta}_{\rm 0}^{2}+4\tau^{2}}}$, 
%where $\tau$ is dimensionless time, appropriately normalized by $\Omega$ and $\theta_{\rm E}$. One sees, that when 
%$\tau \rightarrow 0$, $M(\tau)\rightarrow \frac{1}{\sqrt{\hat{\theta}_{\rm 0}^{2}+\tau^{2}}}$. 
%The lower boundary for $\hat{\theta}_{\rm 0}$ is determined by the angular size of the emitting blob. 
%Knowing $M(\tau)$ one can easily calculate the LC 

Under proper assumptions one can calculate $M(t)$ and respectively -- the light curve
if the emission is dominated by a single, microlensed blob.
In this case the best working fit appears to be the \textit{Lorentzian}, gradually being transformed into an exponent for 
%$\theta_{\rm 0}\rightarrow 0$, 
strong microlensing events, 
see also Paczynski (1996). Interestingly, adding a constant flux as before does not seem to affect significantly 
the shape of the flare (within some limits).

%
%
%\begin{figure}
% \mbox{} \vspace{9.5cm} \special{psfile=fig_d2mn.eps hoffset=-20 voffset=-420
% hscale=60 vscale=60} 
%\caption[]{
%Distributions of the magnitude second derivatives for the blazar sample and the models (see the text).
%}
%\end{figure}

\subsubsection{Comparison with observations}

A dense LC covering a relatively long time interval may allow studying the shapes of the individual flares and thus help 
to constrain the models. As we saw, the latest two models we considered in detail, predict in general different flare shapes, 
which is especially true for the stronger flares. 

%Since the sampling is always an issue for variability studies, we adopt a 
%different approach. 

%Instead of trying to trace individual flares, we compared the observed magnitude change rate 
%($\mathrm{d}m/\mathrm{d}t$ and $\mathrm{d}^{2}m/\mathrm{d}t^{2}$) distributions with the model predictions for the entire sample. Of course, such a comparison 
%can only be qualitative, at least because the time in both models is parameterized by unknown constants. Furthermore, 
%we are fully aware that very different characteristic times may govern different objects or different activity episodes of the same object.
%Comparing the second derivative distributions seems to be more informative as the models predict difference between the 
%positive and the negative parts of the $\mathrm{d}^{2}m/\mathrm{d}t^{2}$ distributions (Fig. 14).

On the other hand, a number of authors have successfully modeled the blazar flares with exponents (Valtaoja et al., 1999; 
B\"{o}ttcher \& Principe, 2009; Hovatta et al., 2009; Chatterjee et al., 2011), 
which resembles the case of a strong microlensing event\footnote{One is to bear in mind, however, that many of these authors used flux units, 
not magnitudes, when fitting the flares}. Most of these authors, however, used different rising and decaying times for their 
exponential fits, which cannot be easily explained in terms of microlensing (see also Reinthal et al., 2012, for statistics on
time-asymmetry for many objects). If not attributed to a possible degeneracy of the fitting procedure when applied to multiple flares, 
such a time asymmetry might be a result from the evolution of the electron energy density distribution, i.e. from different 
characteristic times of the acceleration and energy loss mechanisms.
Having said that, we stress again that to calculate the emission one needs to solve the \textit{Fokker-Planck} equation. 
For example, even if the energy loss is the only mechanism to modify the energy density distribution of the relativistic particles, 
this would not necessarily mean a decreasing flux at certain wavelengths and vice versa\footnote{If due to energy loss, the 
synchrotron flux will drop at all wavelengths only if $\alpha>2$, where $n(\gamma)\propto \gamma^{-\alpha}$}.
Actually, the asymmetric flares can also be a result from jet inhomogeneity with a combination of jet curvature (e.g., Villata et al. 2009).

%Even just qualitative, the comparison between the $\mathrm{d}^{2}m/\mathrm{d}t^{2}$ distributions (Fig. 14) suggests that the variability of 
%the blazar sample can be attributed to a combination of different jet swinging and/or microlensing events, with perhaps 
%different characteristic times for each particular object. 

Unfortunately, during our monitoring we could not trace individual flares, so it is difficult to distinguish which one (if any) 
of the models is mostly responsible to produce the variability (the flares) or even which one plays more important role. 
The situation becomes even more complicated if one considers a combination of different flares with different phases and amplitudes 
to produce the observed variability picture.
%, instead of modeling the $\mathrm{d}^{2}m/\mathrm{d}t^{2}$ distribution from a single flare. 
In any case, the presence of very sharp peaks seems to favor microlensing. It is also possible that different mechanisms 
govern the flux changes at different time scales.

\subsection{Radiation from the black hole vicinity?}

Possible quasi-periodic oscillations with a period of an hour or so (the BL Lac case, Fig. 10), 
if associated with Keplerian motion around the supermassive black hole, may impose certain constraints 
on the black hole characteristics -- the mass ($M_{\rm BH}$) and the spin parameter ($a$). The Keplerian
period (in hours) at a distance $r$ is $t_{\rm K}\simeq (r^{3/2} + a) M_{\rm 8}$, where
$M_{\rm 8}$ is the central mass in 10$^{8}$ Solar masses (e.g., Shapiro \& Teulkolsky 1983),
and $r$ is in units of $M$ (geometrical units used, $G=c=1$). Since the horizon is determined
by $r_{\rm H}=1+(1-a^{2})^{1/2}$, one gets (taking that $t_{\rm K}\simeq 1$) $M_{\rm 8}\leq 0.5$ for
any value of the black hole spin ($0 \leq a\leq 1$). The mass should be even smaller if one 
accepts a much more physical assumption, i.e. that the oscillation is due to some instabilities at the 
innermost stable circular orbit ($r_{\rm ISCO}$), instead of at the horizon. The last distance depends in a complex way on 
the spin and the mass (e.g. Shapiro \& Teulkolsky, 1983). A simple useful fit for practical purposes that works well 
everywhere, except for $a\simeq1$, and combines $t_{\rm K}$ (in hours) at $r_{\rm ISCO}$ with the black hole parameters 
is the following expression: $M_{\rm 8}\simeq t_{\rm K}/\sqrt{330.7~e^{-a}-115.2}$.
Therefore, $M_{\rm BH}$ should be between $7~10^{6}$ $M_{\odot}$ for $a=0$ and $4~10^{7}$ $M_{\odot}$ for $a=0.998$ (Fig. 13)
if the oscillations are due to Keplerian motion at $r_{\rm ISCO}$.
Such a mass is at least an order of magnitude lower than what the host galaxy magnitude suggests for 
BL Lac (Wu et al. 2002; Falomo et al. 2003), but is similar to other estimates, based on the variability
characteristics (Xie et al. 2002). It is clear, that since the link between these possible quasi-periodic oscillations 
and the orbital motion is far from proven, no firm conclusion is possible at this stage. Studying such oscillations, however, 
if furhter confirmed with high-accuracy observations, can become one of the very few ways to probe the blazars' innermost regions.

\begin{figure}
 \mbox{} \vspace{5.0cm} \includegraphics{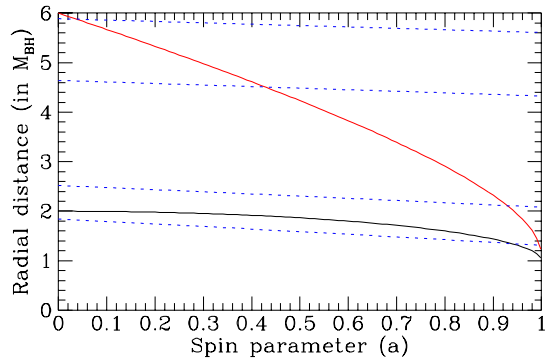} 

\caption[]{The radial distances for an orbital period of 1 hour as functions of black hole mass and 
spin parameter. Thick continuus lines indicate the innermost stable orbit and the horizon, dashed lines --
the 1 hour period distances for $M_{\rm 8}=0.07$, 0.1, 0.25 and 0.4, from top to bottom, respectively.
Note that all the solutions passing below the horizon line are unphysical there.}
\end{figure}

\section{Conclusions}

We publish optical intra-night LCs of 13 blazars (BL Lacs and FSRQs), covering a total of $\sim 160$ hours of monitoring time. 
In the majority of cases we detected no variability at all, and when the variations were found, they resembled mostly linear 
trends or smooth fluctuations. In any case, we observed no erratic, "frame-to-frame" changes, which are sometimes reported. 
The smoothness of the LCs allowed fitting a low-order polynomial, which helped to better reveal the magnitude trends. 

The distribution of the magnitude change rates appears to be time-asymmetric, which, if real, would mostly favor models, based 
on electron energy density evolution rather than "time-symmetric" models like jet swinging or microlensing.

We found some indications for the presence of quasi-periodic micro-oscillations with a period of about an hour in the high 
accuracy LCs of BL Lac after subtracting a leading polynomial.

Unfortunately, at this stage, we were unable to establish which one of the existing models is mostly responsible 
for the blazar variability. It is possible that more than one of them can play 
some role for the different objects and/or activity episodes of the same object. 
Clearly, further high accuracy intra-night variability studies are needed for a firm conclusion.

\section*{Acknowledgments}

This research was partially supported by Scientific Research
Fund of the Bulgarian Ministry of Education and Sciences
(BIn-13/09 and DO 02-85) and by Indo-Bulgarian bilateral
scientific exchange project INT/Bulgaria/B-5/08 funded
by DST, India. We thank the anonymous referee for the rapid response and valuable suggestions. 
This research made use of CDS-Strasbourg (Simbad) and NASA-IPAC (NED) databases.

\bsp

\label{lastpage}

\end{document}